# Single-walled carbon nanotubes and nanocrystalline graphene reduce beam-induced movements in high-resolution electron cryo-microscopy of ice-embedded biological samples

Daniel Rhinow[1*], Nils-Eike Weber[2], Andrey Turchanin[2], Armin Gölzhäuser,[2] Werner Kühlbrandt[1]

[1] *Max-Planck-Institute of Biophysics, Department of Structural Biology, Max-von-Laue-Straße 3, D-60439 Frankfurt, Germany*

[2] *University of Bielefeld, Department of Physics, Universitätsstraße 25, D-33615 Bielefeld, Germany*

**Abstract**

For single particle electron cryo-microscopy (cryoEM), contrast loss due to beam-induced charging and specimen movement is a serious problem, as the thin films of vitreous ice spanning the holes of a holey carbon film are particularly susceptible to beam-induced movement. We demonstrate that the problem is at least partially solved by carbon nanotechnology. Doping ice-embedded samples with single-walled carbon nanotubes (SWNT) in aqueous suspension or adding nanocrystalline graphene supports, obtained by thermal conversion of cross-linked self-assembled biphenyl precursors, significantly reduces contrast loss in high-resolution cryoEM due to the excellent electrical and mechanical properties of SWNTs and graphene.

* corresponding author: daniel.rhinow@biophys.mpg.de, Tel.: +49-69-6303-3050

Electron cryo-microscopy (cryoEM) is a powerful tool in structural biophysics. The structure of biological macromolecules has been analyzed at atomic or near-atomic resolution by cryoEM of vitrified samples.[1,2] While a resolution of 3.3 Å has recently been achieved by single particle cryoEM of aquareovirus,[3] resolutions obtained with most ice-embedded specimens are significantly worse. Although theory indicates that this level of resolution should be attainable routinely,[4] image contrast in high-resolution cryoEM of biological specimens is degraded by radiation damage, inelastic scattering, charging, and beam-induced specimen movement.[5,6] Radiation damage sets an upper limit to the electron dose that can be used for high-resolution imaging of radiation-sensitive biological specimens. At room temperature, this limit is ~1 electron/Å$^2$ but cooling the specimen to 98 K with liquid nitrogen or to 4 K with liquid helium reduces the impact of radiation damage by a factor of 10 or 20, respectively.[7-9]

Numerous efforts have been made to improve instrumentation for transmission electron microscopy (TEM). Energy filters remove inelastically scattered electrons,[10,11] $C_s$ correctors compensate for the spherical aberration of electron lenses,[12,13] and new materials are increasingly used as TEM supports, including conductive amorphous alloys[14] and graphene or graphene-like supports.[15-17] Nonetheless, the most critical factor in high-resolution cryoEM is the specimen itself. For single-particle cryoEM, samples are usually embedded in a thin layer of vitreous water. Vitreous water is an electrical insulator, and therefore cryoEM images of frozen-hydrated specimens suffer from an uncontrolled buildup of electrostatic charge due to beam-induced ejection of electrons. The situation is made worse by the fact that the mechanical strength of a thin film of vitreous water is up to 2 orders lower than that of a carbon film of equal thickness.[6] Thick carbon support films might solve the problem for electron crystallography.[18] However, for single-particle cryoEM, beam-induced specimen movement will remain a serious problem, as thin films of vitreous water spanning the holes of a holey carbon film are particularly susceptible to beam-induced movement and charging.

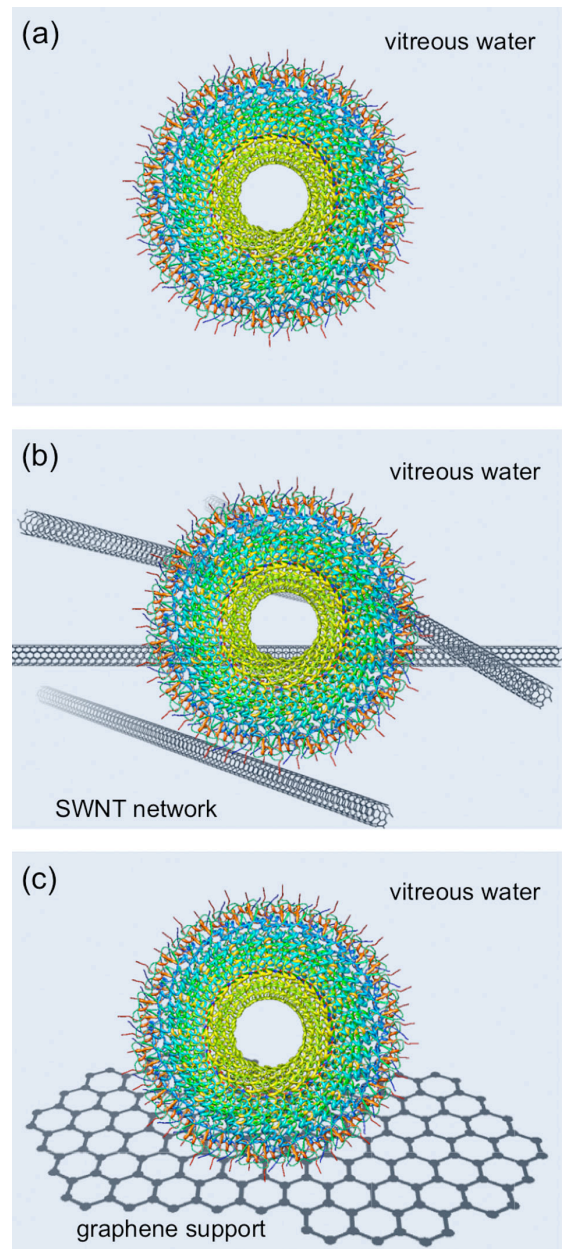

**Figure 1**

Schematic representation of the investigated samples (a) TMV in vitreous ice suspended over holey carbon. (b) TMV in vitreous ice, vitrified together with BSA-solubilized SWNTs, suspended over holey carbon. (c) Ice-embedded TMV on nanocrystalline graphene. Nanocrystalline graphene films are mounted on lacey carbon. Molecular graphics images were produced using CoNTub 2.0 (ref. 31), Chimera (ref. 32), and ArgusLab (M. A. Thompson, Planaria Software LLC, Seattle, WA, http://www.arguslab.com).

In this letter we show that these problems can be at least partly overcome by doping the cryoEM sample with single-walled carbon nanotubes (SWNT) or adding ultrathin supports comprising nanocrystalline graphene. SWNTs are electrically conducting quantum wires with unique mechanical properties, which make them promising building blocks for nanotechnology.[19-21] The ability of SWNTs to dissipate electrostatic charges effectively, along with their exceptional mechanical strength predestines them for cryoEM specimen preparation. So far, the use of SWNTs in biological applications has been largely precluded by their extreme hydrophobicity. Recently a variety of dispersing agents have been tested, which solubilize SWNTs in water, among them the protein bovine serum albumin (BSA).[22,23]

We used tobacco mosaic virus (TMV) as a test specimen to show that contrast loss with cryoEM specimens, vitrified together with BSA-solubilized SWNTs (Fig. 1b), is significantly decreased compared to the same specimens embedded in vitreous water only (Fig. 1a). Furthermore, we compare vitrification with SWNTs with vitrification of biological samples on graphene supports (Fig. 1c), made by thermal conversion of cross-linked aromatic self-assembled monolayers[24] into free-standing films comprising nanocrystalline graphene.[25,26] TMV was a kind gift of Carsten Sachse (EMBL, Heidelberg, Germany). All chemicals were purchased from Sigma-Aldrich Inc. and used as received. SWNTs were solubilized with BSA as dispersing agent.[23] 1 mg SWNTs (0.7-1.3 nm diameter) were suspended in 1 mg/ml BSA solution and sonicated for 30 min, resulting in a stable and homogenous, black aqueous suspension, which was analyzed by cryoEM. Aqueous solutions of TMV were mixed with BSA-solubilized SWNTs at 1:1 ratio and were applied to holey carbon (Quantifoil) grids glow discharged in air for 60 s. Alternatively, TMV was applied to nanocrystalline graphene, mounted on lacey carbon grids, glow discharged in air for 30 s. Samples were blotted for 2 sec and plunged in liquid ethane using a FEI Vitrobot III.

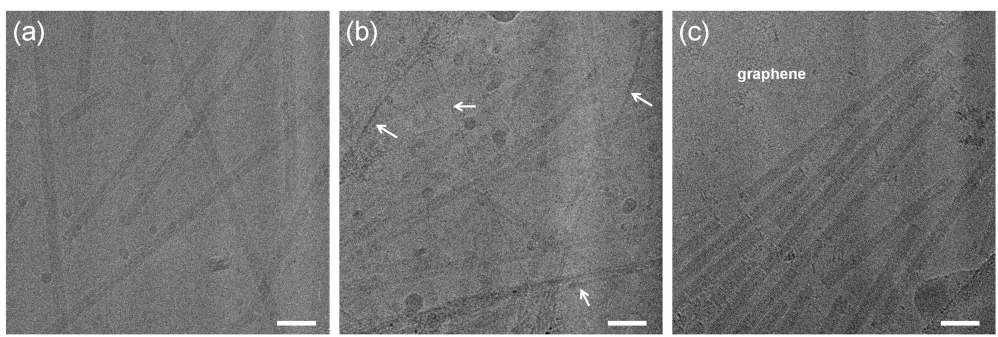

**Figure 2**

CryoEM of TMV.(a) TMV in vitreous ice at 4 K imaged in a JEOL-3000 SFF microscope. (b) TMV/SWNTs in vitreous ice at 4 K. Shown is a digitized micrograph, which reveals individual SWNT fibers (arrows) along with TMV particles. (b) Ice-embedded TMV on nanocrystalline graphene support. Scale bars are 50 nm.

For quantitative analysis of image contrast, vitrified samples were analyzed at 4 K in a liquid helium cooled JEOL-3000 SFF microscope at 300 kV. At 4 K conventional amorphous carbon is virtually an electrical insulator, making effects of specimen charging and beam-induced movement particularly noticeable. The top-entry specimen holder of the microscope minimizes drift as a source of contrast loss. Figure 2a shows ice-embedded TMV imaged at 4 K. A digitized micrograph of ice-embedded TMV/SWNTs is shown in Figure 2b. Individual SWNT fibers (arrows) form networks in the vitrified water along with TMV particles. Figure 2c shows a digitized micrograph of ice-embedded TMV on a graphene support recorded at 4 K. It is worth noting that ultrathin nanocrystalline graphene supports withstand plunging forces routinely and form stable supports at 4 K.

Image contrast of 18 individual TMV particles on the best micrographs (6 of TMV, 6 of TMV/SWNT, and 6 of TMV on nanocrystalline graphene out of 20 micrographs in each case) was evaluated by calculating the intensity ratio of the 3$^{rd}$ and 6$^{th}$ layer line in Fourier space.[6,27] This ratio includes information both about the molecular envelope (3$^{rd}$ layer line) and the molecular interior (6$^{th}$ layer line) of TMV. Any loss of image contrast due to charging and beam-induced movement weakens the high-resolution information, and thus increases the ratio of $3^{rd}/6^{th}$ layer line intensities. Figure 3a-c shows Fourier transforms of representative images together with line scans perpendicular to the direction of the layer lines. The $3^{rd}$ and $6^{th}$ layer lines of TMV were visible in all images, indicating a resolution of 23 Å and 11.5 Å respectively. The ratio of $3^{rd}/6^{th}$ layer line intensities was calculated for all 18 TMV particles using imageJ (Rasband, W.S., U. S. National Institutes of Health, Bethesda, Maryland, USA, http://rsb.info.nih.gov/ij/, 1997-2009) and is given in Figure 3d. In each case the image contrast of ice-embedded TMV vitrified together with SWNTs was better than that of ice-embedded TMV without SWNTs by a factor of 1.57 on average. Comparable contrast improvement was obtained with TMV on nanocrystalline graphene supports, where the factor was 1.5 compared to TMV without graphene or SWNTs. The observed improvement in image contrast did not depend on the contact of an individual virus particle with a SWNT, nor its position relative to the edge of the Quantifoil film. Our results indicate that the excellent electrical and mechanical properties of SWNTs and graphene reduce beam-induced movements, thus improving image contrast of ice-embedded biological specimens.

We have demonstrated that ice-embedded biological samples show significantly enhanced image contrast when prepared together with SWNTs in a thin film of vitreous water on holey carbon film. The gain in image contrast obtained with TMV vitrified together with SWNTs resembled the results obtained with TMV vitrified on nanocrystalline graphene. Adding SWNT networks or graphene supports may be the simplest and best method presently available for stabilizing vitrified specimens for single-particle cryoEM. SWNTs are easily available and inexpensive in the small amounts needed for cryoEM specimen preparation. Furthermore, a variety of methods for fabricating freestanding carbon nanotube thin films has been developed recently, which could be adopted to pre-print SWNT networks on holey carbon TEM grids prior to sample preparation, thus superseding the use of dispersing agents.[28-30]

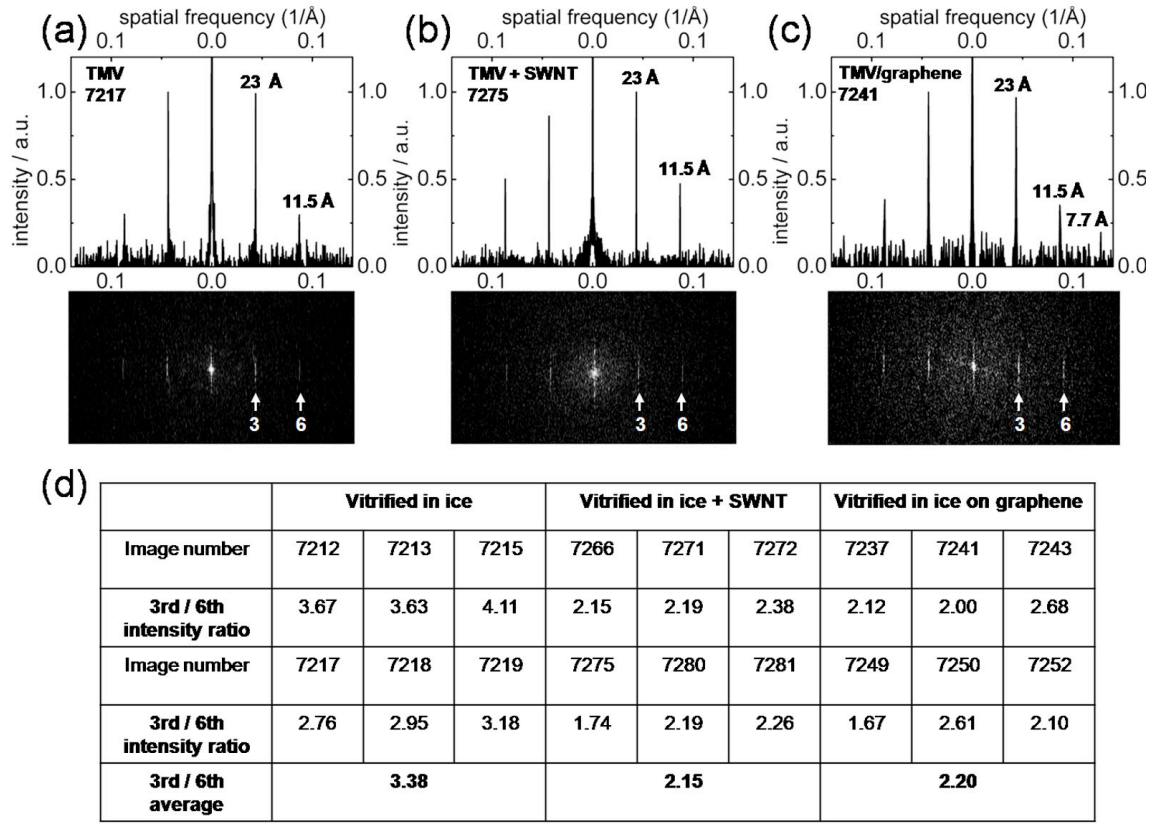

**Figure 3**

Image contrast of ice-embedded TMV. Shown are Fourier transforms of individual TMV particles along with line sections perpendicular to the layer lines (normalized to the 3$^{rd}$ layer line). Arrows in the Fourier transforms indicate the 3$^{rd}$ and 6$^{th}$ layer lines. Fourier transforms were generated using the 2dx software (ref. 33), which is based on the MRC package (ref. 34). (a) TMV in ice. (b) TMV in ice, vitrified together with SWNTs. (c) Ice-embedded TMV on graphene. The 9$^{th}$ layer line indicates strong contrast at 7.7 Å resolution. (d) Quantitative analysis of image contrast. The ratio of 3$^{rd}$/6$^{th}$ layer line intensity calculated for the respective layer lines indicates clearly that SWNTs or graphene reduce contrast loss.